\documentclass[prc,twocolumn,floatfix,showpacs,preprintnumbers,amsmath,amssymb]{revtex4}

\usepackage{graphicx}
\usepackage{dcolumn}    
\usepackage{bm}         
\def \beq {\begin{equation}}
\def \eeq {\end{equation}}
\def \beqa {\begin{eqnarray}}
\def \eeqa {\end{eqnarray}}

\newcommand{\vect}[1]              
           {\mbox{\boldmath$#1$}}  
\begin{document}
%

\title{Electron scattering in isotonic chains as a probe of the proton shell
structure of unstable nuclei}

\author{X. Roca-Maza\textsuperscript{1,2,3}}
\email{xavier.roca.maza@mi.infn.it}
\author{M. Centelles\textsuperscript{1}}
\author{F. Salvat\textsuperscript{1}}
\author{X. Vi\~nas\textsuperscript{1}}

\affiliation{\textsuperscript{1} Departament d'Estructura i Constituents de la Mat\`eria
             and Institut de Ci\`encies del Cosmos, Facultat de F\'{\i}sica, Universitat 
             de Barcelona, Diagonal {\sl 645}, {\sl 08028} Barcelona, Spain\\
             \textsuperscript{2} Dipartimento di Fisica, Universit\`a degli Studi di Milano, 
             via Celoria 16, {\sl I-20133} Milano, Italy\\
             \textsuperscript{3} INFN, sezione di Milano, via Celoria 16, {\sl I-20133}
             Milano, Italy}


%
\begin{abstract}
Electron scattering on unstable nuclei is planned in future facilities
of the GSI and RIKEN upgrades. Motivated by this fact, we study
theoretical predictions for elastic electron scattering in the
$N=82$, $N=50$, and $N=14$ isotonic chains from very proton-deficient to very
proton-rich isotones. We compute the scattering observables by performing Dirac
partial-wave calculations. The charge density of the nucleus is obtained
with a covariant nuclear mean-field model that accounts for the low-energy
electromagnetic structure of the nucleon. For the discussion of the dependence
of scattering observables at low-momentum transfer on the gross properties of
the charge density, we fit Helm model distributions to the self-consistent
mean-field densities. We find that the changes shown by the electric charge form
factor along each isotonic chain are strongly correlated with the underlying
proton shell structure of the isotones. We conclude that elastic electron scattering experiments in isotones can provide valuable information about the filling order and occupation of the single-particle levels of protons.
\end{abstract}

\maketitle

%
\section{Introduction}

Since the 1950's, elastic electron scattering has been utilized to obtain
accurate information on the size and shape of nuclei
\cite{Hof56,Don75,Don84,Moya86,Sick01}. Because electrons and nucleons interact
essentially through the electromagnetic force, the nucleus remains
rather unperturbed during the scattering process and the analysis of the data
is not hampered by uncertainties associated with the strong interaction.
Thus, electron scattering is able to provide
very clean information about the charge distribution of atomic
nuclei~\cite{Vri87,Fri95,Ang04}.

Low-energy nuclear physics is nowadays moving very fast towards the domain of
exotic nuclei \cite{ENAM08}. This is due to the development of successive
generations of radioactive-isotope beam (RIB) facilities
\cite{Tan95,Gei95,Mue01,sud11,xia02,xia07}, such as FAIR and SPIRAL2 in Europe,
FRIB in North America, and HIRFL-CSR, RARF or RIBF in Asia, which will allow
studying the properties of nuclei beyond the stability valley. Many interesting
effects have already been discovered in exotic nuclei, such as neutron and
proton halos, neutron skins, and new magic numbers. These effects may be related
to the structure of the nucleon distributions far from stability. As
with stable nuclei, one way of exploring the structure of exotic nuclei is
through the electromagnetic interaction. For this purpose, a new generation of
electron-RIB colliders using storage rings is under construction by RIKEN
(Japan) \cite{sud11,Kat03} and at GSI (Germany) \cite{GSI02,Sim04}. It is expected
that in the near future the SCRIT project in Japan \cite{Sud05,Wak08,Sud09}
and the ELISe experiment at FAIR in Germany \cite{Sim07,Ant11} will offer the
opportunity of studying the structure of unstable exotic nuclei by means of
electron scattering.

On the theoretical side, much work has been devoted to the study of charge
distributions of exotic nuclei through calculations of both electron scattering
(see e.g.\ Refs.\ \cite{Gar99,Ant05,Sar07,Karat07,Bertu07,Zai04,Zai06,Roc08})
and proton scattering (see e.g.\ Ref.\ \cite{Amos04}). Suda \cite{Sud04} pointed out that in electron scattering off unstable nuclei the maxima and the minima of the charge form factor are very sensitive to the size and the diffuseness of the charge density. This fact has been confirmed by different works that have analyzed the behavior of the charge form factor along
isotopic \cite{Ant05,Sar07,Zai04,Roc08} and isotonic \cite{Zai06} chains.

To probe the charge distribution in nuclei, the electron beam energy needs to be
of the order of a few hundred MeV. As one deals with relativistic electrons, it
is mandatory to solve the elastic scattering problem of Dirac particles in the
potential generated by the nuclear charge density. The simplest approach is the
plane-wave Born approximation (PWBA) where the initial and final states of the
electron are described by Dirac plane waves. The PWBA accounts for many
features of electron scattering but it cannot provide an accurate description of
the electric charge form factor, in particular near the dips. The most
elaborated calculations of electron-nucleus scattering are 
obtained by the exact phase-shift analysis of the Dirac equation.
This calculation scheme is known as distorted-wave Born approximation
(DWBA) \cite{Yen54}. 
The DWBA has been used to analyze different
aspects of the scattering of electrons by nuclei, see e.g.\ Refs.\
\cite{Heisenberg81,Nishimura85,Udias93,Caballero98,Ant05,Sar07,Roc08,Chu10} 
and references therein. 
In the present work we employ the DWBA to study elastic
electron scattering in isotones.
It may be mentioned that the eikonal approximation has been
applied in some studies of elastic electron scattering off
proton-rich and neutron-rich nuclei \cite{Zai04,Zai06,Chu09}.

The charge density of the target nucleus is one of the basic ingredients of the electron-nucleus scattering problem. For medium and heavy nuclei, the theoretical charge densities can be calculated in the mean-field approximation using non-relativistic nuclear forces or relativistic mean field (RMF) models.
It is known that the overall trends of the elastic scattering of electrons by
stable medium and heavy nuclei, are well reproduced by the mean-field
charge densities computed with nuclear models that have been calibrated to
describe the ground-state properties (in particular the charge radii) of some
selected nuclei. However, different nuclear models differ in the fine
details and describe with different quality the experimental scattering data.
See Ref.~\cite{Roc08} for a recent comparison of the elastic electron scattering results predicted by different nuclear mean-field models.

In Ref.~\cite{Roc08} we studied elastic electron scattering along the
Ca and Sn isotopic chains in DWBA. In that work we reported
several correlations among scattering observables and some properties of the
nuclear charge density along the isotopic chains \cite{Roc08}. In the present work, we investigate what information on nuclear structure can be gained
from the study of elastic electron scattering in the $N=82$, 50, and 14
isotonic chains. We aim at extracting general trends, according to current mean-field theories, about the behavior of some observables that may be measured in experiments performed with unstable nuclei in the low-momentum transfer region.
Our choice of the $N=82$, 50, and 14 isotonic chains among other possible $N$
values, is mainly motivated by the fact that they cover different
regions of the mass table and by the following reasons. On the one hand, there
is a certain interest in the structure of unstable nuclei belonging to the
$N=82$ and $N=50$ shell closures because some of these nuclei may correspond to
waiting points in the astrophysical $r$-processes of nucleosynthesis
\cite{arno07,paar07,Grawe07}.
The $N=82$ isotones below $^{132}$Sn are believed to be in close relation with the peak of the solar $r$-process abundance distribution observed around the mass number $A=130$ \cite{Grawe07,Jungclaus07}, whereas the $N=50$ isotones near $^{78}$Ni are thought to be responsible for producing the pronounced abundance peak observed around $A=80$ \cite{Grawe07,Baruah08}.
On the other hand, scattering data for light nuclei, such as e.g.\
those of $N=14$, are likely to be obtained in future electron scattering
facilities such as SCRIT \cite{Sud05,Wak08,Sud09} and ELISe \cite{Sim07,Ant11}.

The study of elastic electron scattering along isotopic
and isotonic chains explores different aspects of the nuclear charge density. The
electric charge form factor along an isotopic chain gives information about
the effect of the different number of neutrons on the charge density, which
becomes more and more dilute and extends to larger distances as the neutron
number increases \cite{Roc08}. In an isotonic chain, the changes in the charge
form factor primarily inform about the effect of the outer proton single-particle
orbitals that are being filled as the atomic number increases in the chain.
Thus, our previous \cite{Roc08} and
present study together provide a survey of the evolution of the charge form factor with
the neutron and proton numbers in different mass regions of the nuclear chart.

The rest of this article is organized as follows. In Section II, we summarize
the method employed in our study of electron scattering in isotonic chains. As
the basic methodology follows that of Ref.~\cite{Roc08}, we address the reader
to that work and references therein for more details about the relativistic
nuclear mean-field theory and about the Dirac partial-wave analysis, which we
perform using the ELSEPA code \cite{Sal05} adapted to the nuclear problem. We
devote Section III to the presentation and analysis of our numerical results for
elastic electron-nucleus scattering in the $N=82$, $50$, and $14$ chains.
Finally, our conclusions are laid in Section IV.

\section{Method}
\label{method}

To investigate electron scattering in isotonic chains we follow the method developed in 
Ref.~\cite{Roc08}. For completeness, we summarize here the main aspects of this method. 
The electron beam energy in our investigation is fixed at 500 MeV, which is a typical energy 
in electron-nucleus scattering experiments. Indeed, rather than discussing directly the
differential cross section (DCS), we study the DWBA electric charge form factor because it 
is almost independent of the electron beam energy in the low-momentum transfer regime, 
as it can be seen from Fig.~5 of Ref.~\cite{Roc08} and from Fig.~\ref{dcs-ff82}.b below.
We compute the electric charge form factor as follows \cite{Roc08}:
\begin{equation}
{\vert F(q) \vert}^2 =
\Big( \frac{d \sigma}{d \Omega} \Big)
\Big/ \Big( \frac{d \sigma_{\rm point}}{d \Omega} \Big),
\label{fdwba}
\end{equation}
where $d\sigma/d \Omega$ and $d\sigma_{\rm point}/d \Omega$ are the DCS of the
extended nucleus and of the point nucleus, respectively, calculated in DWBA.
We denote the form factor (\ref{fdwba}) by $F_{\rm DWBA}(q)$ hereinafter.
It is to be mentioned that here we are using the DWBA point-nucleus DCS rather than the usual Mott cross section \cite{Preston82}:
\begin{equation}
\frac{d \sigma_{\rm Mott}}{d \Omega} =
\bigg(\frac{Ze^2}{2E}\bigg)^2 \frac{\cos^2(\theta/2)}{\sin^4(\theta/2)} .
\label{sigmamott}
\end{equation}
In order to extract the effect of the finite size of the nucleus it seems reasonable to consider the two cross sections in Eq.~(\ref{fdwba}) calculated within the same approximation. The point-nucleus DCS calculated within the DWBA
was also used in analyses of the form factor of elastic electron scattering data (see e.g.\ Refs.\ \cite{Hof53,Helm56} and discussions in Ref.~\cite{Schiff53}). We make a comparison of the results of the two approaches in Section III below.
It is found that the choice is not critical for our study in the low-momentum transfer regime.

We calculate the charge densities with the relativistic
mean-field parametrization G2 \cite{G2,Ser97}, which we also employed in
Ref.~\cite{Roc08}. This nuclear model was
constructed as an effective hadronic Lagrangian consistent with the symmetries
of quantum chromodynamics. The nucleon density distributions are obtained
self-consistently at the mean-field level by numerical solution of the
corresponding variational equations. In contrast to most of the nuclear
mean-field models that assume point densities, the G2 Lagrangian
incorporates the low-energy electromagnetic structure of the nucleon through
vector-meson dominance \cite{G2,Ser97}. This implies that the charge density
is obtained directly from the self-consistent solution of the mean-field
equations without any extra folding with external single-nucleon form factors.
We have verified that the charge density distribution provided by G2
agrees very well with the charge density that can be obtained from the point proton and point neutron density distributions of G2 folded with experimental
single-nucleon charge form factors.
It has been shown \cite{Ser97,G2,Arumugam04} that the G2 relativistic mean-field
interaction is a reliable parameter set both for calculations of ground-state
properties of nuclei and for predictions of the nuclear equation of state up to
supra-normal densities, as well as for predictions of some properties of neutron
stars. First calculations of the charge form factor in PWBA with G2 were
reported in Ref.\ \cite{G2}.

In our calculations we assume spherical symmetry although some of
the nuclei considered in this study may be deformed, particularly in
the case of the $N=14$ isotones \cite{Doba04,Lala99}.
Pairing correlations are important for describing
open-shell nuclei. We take pairing into account through a modified BCS approach
that simulates the continuum (needed for nuclei at the drip lines) through
quasi-bound levels which are retained by their centrifugal barrier (neutron
levels) or by their centrifugal-plus-Coulomb barrier (proton levels)
\cite{Est01b}. The pairing interaction in this approach is described by means of
a constant matrix element fitted to reproduce the experimental binding energies
of some selected isotopic and isotonic chains as described in Ref.\
\cite{Est01b}. It is to be mentioned that a mean-field treatment is not
expected to be sufficient for light exotic nuclei~\cite{Karat07}. Thus, the
$N=14$ isotonic chain
studied below (with nuclei from $^{22}$O to $^{34}$Ca) corresponds to a somewhat
limiting case, and the mean-field results should be taken as
semiquantitative. The calculations with the G2 model predict a relatively
magic character of the neutron numbers $N=14$ and $N=16$.
These neutron numbers have attracted some attention in recent theoretical and
experimental studies as possible new magic numbers in exotic nuclei
\cite{Ozawa00,Otsuka01,Gupta06,Tanaka09,Wang10}.

The use of modeled charge densities and electric charge form factors in the
experimental analysis of scattering data has been extensive in the past, and
continues to date. This is because in many cases the parameters of the
modeled charge densities are directly related with the size of the bulk and
surface regions of the nucleus under study. In this way, the modeled
densities help to provide a clear physical interpretation of the electron
scattering data. This is the case of the so-called Helm model \cite{Helm56} that
we used for some calculations in our previous study of isotopic chains
\cite{Roc08}. The parameters of the Helm model are fitted to the electric
charge form factor in the low-momentum transfer regime. Here, the calculations
with the Helm model will be helpful to gain some insight about the variation of the position and width of the surface of the charge density distribution along the isotonic chains. We briefly summarize the fitting procedure of the parameters of the Helm model in the next subsection.

\subsection{Equivalent Helm charge densities}
\label{Sect-Helm}

The original version of the Helm model \cite{Helm56} has been extended in
various ways for a more accurate description of the experimental charge
densities \cite{Frie82,Frie86,Spru92}. In the simplest version of the model
\cite{Helm56}, the charge density is obtained from the convolution of a constant
density $\rho_0$ in a hard sphere of radius $R_0$ with a Gaussian distribution
having variance $\sigma^2$. By construction, $R_0$ gives the effective location
of the position of the nuclear surface, and hence characterizes the size of the
density profile, whereas the parameter $\sigma$ is a measure of the thickness of
the surface region of the density distribution. The Helm charge density is then
given by
\begin{equation}
\rho^{(H)}(\vec{r}) = \int d \vec{r}' f_G (\vec{r} -\vec{r}') \rho_0
\Theta(R_0 - r),
\label{eq14}
\end{equation}
where
\begin{equation}
f_G(r) = \big( 2 \pi \sigma^2 \big)^{-3/2} e^{-r^2/2 \sigma^2}.
\label{eq15}
\end{equation}
The two parameters, $R_0$ and $\sigma^2$, of the Helm model determine the charge density as well as the electric charge form factor within the PWBA:
\begin{equation}
F^{(H)}(q) = \int e^{i \vec{q}\cdot\vec{r}} \rho^{(H)}(\vec{r})
d\vec{r} = \frac{3}{q R_0} j_1(q R_0) e^{- \sigma^2 q^2/2} ,
\label{eq17}
\end{equation}
where $j_1 (x)$ is the spherical Bessel function. Note that we use natural
units throughout the present paper.

We proceed as suggested originally in Ref.~\cite{Helm56} to obtain
the Helm parameters associated to a given nucleus from the PWBA electric
charge form factor of that nucleus. First, we require that the first zero of
Eq.\ (\ref{eq17}) coincides with the first zero of the mean-field PWBA charge
form factor (Fourier transform of the charge density obtained with the G2
model). We will refer to this charge form factor as $F_{\rm PWBA}(q)$
hereinafter. Therefore, the radius of the equivalent Helm density reads
\begin{equation}
R_0 = \frac{x}{q_0}.
\label{eq18}
\end{equation}
where $x=4.49341$ is the first zero of $j_1(x)$ and $q_0$ is the momentum transfer 
corresponding to the first zero of $F_{\rm PWBA}(q)$. Second, we determine the 
variance $\sigma^2$ of the Gaussian distribution such that 
$\vert F^{(H)}(q_{\rm max})\vert = \vert F_{\rm PWBA}(q_{\rm max}) \vert$, 
where $q_{\rm max}$ is the momentum transfer corresponding to the second maximum of  
$\vert F_{\rm PWBA}(q) \vert$ (the first maximum appears always at $q=0$ fm$^{-1}$). Using 
Eq.\ (\ref{eq17}), one easily 
obtains
\begin{equation}
\sigma^2 = \frac{2}{q_{\rm max}^2} 
\ln \bigg( \frac{3 j_1(q_{\rm max}R_0)}{
q_{\rm max}R_0 F_{\rm PWBA}(q_{\rm max})} \bigg) .
\label{eq19}
\end{equation}

\section{Results: $N = 82$, $N=50$, and $N=14$ isotonic chains}

We start with the discussion of the results for the $N=82$ isotonic chain where the different aspects of our study are described in detail. After that, we extend our study to the $N=50$ and $N=14$ isotonic chains.

\subsection{$N=82$ isotonic chain}

\begin{figure}[t]
\begin{center}
\includegraphics[width=0.95\linewidth,angle=0,clip=true]{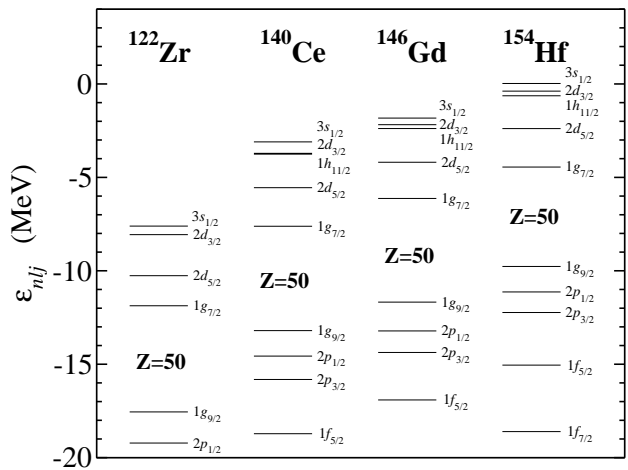}
\caption{\label{spln82} Energy of the proton single-particle levels for $^{122}_{40}$Zr, $^{140}_{58}$Ce, $^{146}_{64}$Gd and $^{154}_{72}$Hf as computed with the relativistic nuclear mean field interaction G2.}
\end{center}
\end{figure}

We first analyze the charge densities along the $N=82$ chain. The ordering and the energy of the different proton single-particle levels, mainly the levels closest to the Fermi level, are quite important for the present study. This is because the corresponding single-particle wave functions determine, to a large extent, the shape of the charge density at the surface region as well as the electric charge form factor in the low-momentum transfer region. Fig.~\ref{spln82} displays the energy of the proton single-particle levels of some selected nuclei of the $N=82$ isotopic chain. They are representative of proton deficient nuclei ($^{122}_{40}$Zr), stable nuclei ($^{140}_{58}$Ce), proton rich-nuclei ($^{146}_{64}$Gd) and proton drip-line nuclei ($^{154}_{72}$Hf).

The more relevant proton single-particle levels in our analysis of the $N=82$ isotonic chain are, on the one hand, the $1g_{9/2}$, $1g_{7/2}$ and $2d_{5/2}$ levels (which appear clearly separated in energy) and, on the other hand, the nearly degenerate $1h_{11/2}$, $2d_{3/2}$ and $3s_{1/2}$ levels (which have a very close energy). The 1$h_{11/2}$ level shows energy gaps of about 2 and 4 MeV with respect to the $2d_{5/2}$ and $1g_{7/2}$ levels, respectively, and a gap of about 9 MeV with respect to the deeper 1$g_{9/2}$ level. This large energy gap is due to the magicity of the proton number $Z=50$. With increasing mass number these relevant levels are shifted up in energy, roughly as a whole, retaining the same ordering and approximately the same energy gaps. As a consequence of this level scheme, in going from the nucleus $^{122}_{40}$Zr to $^{140}_{58}$Ce the charge densities differ basically by the effects of filling up the $1g_{9/2}$ and $1g_{7/2}$ shells, and in going from $^{140}_{58}$Ce to $^{146}_{64}$Gd the charge densities differ by the occupancy the $2d_{5/2}$ shell. In these proton-rich isotones with mass number above $A=140$, the pairing correlations play a non-negligible role and therefore the charge densities also get contributions from the $1h_{11/2}$, $2d_{3/2}$, and $3s_{1/2}$ orbitals. Finally, in the case of the drip-line nucleus $^{154}_{72}$Hf, all of the mentioned single-particle wave functions contribute significantly to the charge density.
\begin{figure}[t]
\begin{center}
\includegraphics[width=0.95\linewidth,angle=0,clip=true]{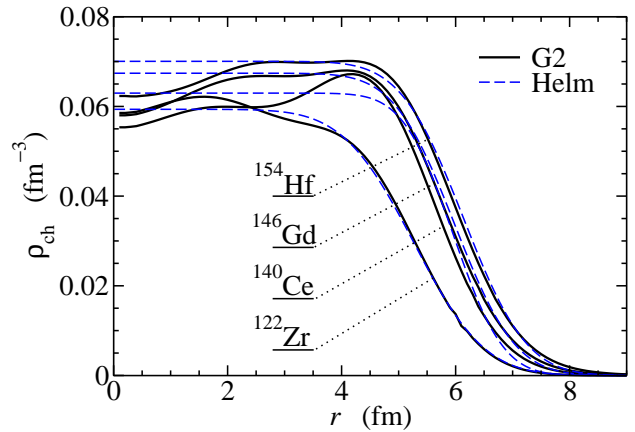}
\caption{\label{dchn82} (Color online) Charge densities for $^{122}_{40}$Zr, $^{140}_{58}$Ce, $^{146}_{64}$Gd and $^{154}_{72}$Hf as a function of the radial distance to the center of the nucleus according to the covariant model G2 (solid lines) and to the fitted Helm distributions (dashed lines).}
\end{center}
\end{figure}
The differences in the charge distribution due to single-particle effects become
evident in Fig.~\ref{dchn82} where the charge densities of
$^{122}_{40}$Zr,$^{140}_{58}$Ce, $^{146}_{64}$Gd, and $^{154}_{72}$Hf computed
with the relativistic mean field model G2 are displayed as functions of the
radial distance.

The equivalent Helm charge densities of these isotones, with
parameters determined as explained in Section \ref{Sect-Helm}, are depicted in
Fig.~\ref{dchn82} by dashed lines. As in the case of isotopes studied in
Ref.~\cite{Roc08}, the quantal oscillations of the mean-field charge densities
are nicely averaged by the bulk part of the Helm model densities. In spite of
the fact that the surface fall-off of the Helm densities is of Gaussian type,
the agreement at the surface between the mean-field and the equivalent Helm
charge distributions is in general satisfactory.
We are aware that a better reproduction of the charge density can be
achieved by using an extended Helm model fitted up to larger values of the
momentum transfer \cite{Frie82,Frie86,Spru92}. However, here we restrict
ourselves to the two-parameter Helm model introduced in Sec.~II.A by the following reasons. On the one hand, the low-momentum transfer region of the electric charge form factor relevant for our study is well enough reproduced by this simple Helm model. On the other hand, this model can provide some transparent information about two main global properties of the underlying charge distribution, namely, its size and surface diffuseness.

\begin{figure}[t]
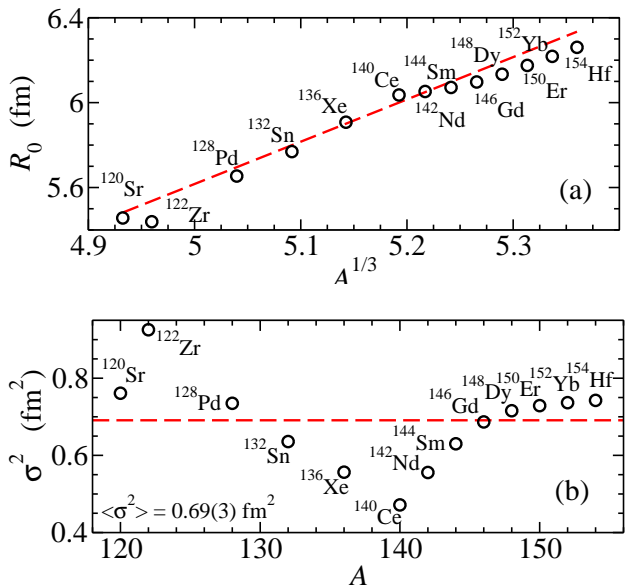

\begin{center}
\includegraphics[width=0.95\linewidth,angle=0,clip=true]{fig03a}
\vspace*{3mm}

\includegraphics[width=0.95\linewidth,angle=0,clip=true]{fig03b}
\caption{\label{r0n82} 
(Color online) Mass-number dependence of the Helm parameter $R_{0}$ predicted by the covariant mean field model G2 in the $N=82$ isotonic chain (a). Mass-number dependence of the Helm parameter $\sigma^2$ (b). The average value is depicted by a horizontal dashed line.}
\end{center}
\end{figure} 

The fitted parameters of the equivalent Helm densities along the $N=82$ isotonic
chain, namely $R_0$ and $\sigma^2$, are given in Table \ref{r-sp-table}. In the
two panels of Fig.~\ref{r0n82} we display these two parameters as function of
$A^{1/3}$ and $A$, respectively. The radius $R_0$ roughly follows a linear trend
with $A^{1/3}$ as it can be expected from the increase of the total number of
nucleons. The parameter $\sigma^2$, which determines the surface thickness of
the charge density, shows a non-uniform variation with the mass number $A$,
caused by the underlying shell structure of the nuclei of this chain.
\begin{table*}[t]
\begin{center}
\caption{Helm model parameters $R_0$ and $\sigma^2$ for the studied isotonic chains of $N=82$, $N=50$ and $N=14$.}
\begin{tabular}{lccclccclccc}   
\hline\hline
             & $N=82$  &          & &             & $N=50$  &          & &             & $N=14$  &          \\
             &         &          & &             &         &          & &             &         &          \\
 Nucl.       &   $R_0$ &$\sigma^2$& &  Nucl.      &   $R_0$ &$\sigma^2$& & Nucl.        &   $R_0$ &$\sigma^2$\\
             &    (fm) &  (fm$^2$)& &             &    (fm) &  (fm$^2$)& &             &    (fm) &  (fm$^2$)\\
\hline
             &         &          & &             &         &          & &             &         &          \\
${}^{120}$Sr  & 5.46    & 0.761    & & ${}^{70}$Ca  & 4.34    & 0.847    & & ${}^{22}$O  & 2.89    & 0.657    \\
${}^{122}$Zr  & 5.44    & 0.926    & & ${}^{74}$Cr  & 4.58    & 0.693    & & ${}^{24}$Ne & 3.07    & 0.689    \\
${}^{128}$Pd  & 5.65    & 0.735    & & ${}^{78}$Ni  & 4.76    & 0.538    & & ${}^{26}$Mg & 3.23    & 0.677    \\
${}^{132}$Sn  & 5.77    & 0.636    & & ${}^{80}$Zn  & 4.85    & 0.533    & & ${}^{28}$Si & 3.36    & 0.680    \\
${}^{136}$Xe  & 5.91    & 0.556    & & ${}^{82}$Ge  & 4.92    & 0.524    & & ${}^{30}$S  & 3.41    & 0.891    \\
${}^{140}$Ce  & 6.04    & 0.472    & & ${}^{84}$Se  & 4.98    & 0.540    & & ${}^{32}$Ar & 3.55    & 0.961    \\
${}^{142}$Nd  & 6.05    & 0.556    & & ${}^{86}$Kr  & 5.01    & 0.632    & & ${}^{34}$Ca & 3.69    & 0.988    \\
${}^{144}$Sm  & 6.07    & 0.630    & & ${}^{88}$Sr  & 5.02    & 0.754    & &             &         &          \\
${}^{146}$Gd  & 6.10    & 0.687    & & ${}^{90}$Zr  & 5.05    & 0.836    & &             &         &          \\
${}^{148}$Dy  & 6.13    & 0.716    & & ${}^{92}$Mo  & 5.12    & 0.830    & &             &         &          \\
${}^{150}$Er  & 6.18    & 0.729    & & ${}^{94}$Ru  & 5.19    & 0.814    & &             &         &          \\
${}^{152}$Yb  & 6.22    & 0.737    & & ${}^{96}$Pd  & 5.25    & 0.790    & &             &         &          \\
${}^{154}$Hf  & 6.26    & 0.743    & & ${}^{98}$Cd  & 5.32    & 0.761    & &             &         &          \\
             &         &          & & ${}^{100}$Sn & 5.39    & 0.727    & &             &         &          \\
             &         &          & &             &         &          & &             &         &          \\
\hline\hline                                                                                                   
\end{tabular}                                                                                                  
\label{r-sp-table}                                                                                             
\end{center}
\end{table*}

In our study of isotopic chains \cite{Roc08}, we found a similar behavior of the
$\sigma^2$ parameter but with two important differences. First, the range of
variation of $\sigma^2$ in isotopic chains is much smaller than the one
exhibited by the $N=82$ isotonic chain. Second, in the case of the Sn isotopes
(see Fig. 6 of Ref. \cite{Roc08}) $\sigma^2$ displays local minima for
$^{132}$Sn and $^{176}$Sn, pointing out the magicity of the $N=82$ and $N=126$
neutron numbers which makes the charge densities of these isotopes more compact.
In contrast, in the $N=82$ isotonic chain, the kinks shown by $\sigma^2$ are rather related with the filling of the different proton single-particle orbitals belonging to the major shell between $Z=50$ and $Z=82$. In particular, when the $1g_{9/2}$ and $1g_{7/2}$ shells are being filled, i.e., between $^{122}_{40}$Zr and  $^{140}_{58}$Ce, $\sigma^2$ decreases almost linearly. The
local minimum of $\sigma^2$ for $^{140}_{58}$Ce points to some magic character
of this nucleus. The fact that the $\sigma$ parameter takes the smaller
values in the region around $^{140}_{58}$Ce, indicates that the surface of the equivalent charge density is more abrupt at and around this
nucleus. When the $2d_{5/2}$ level starts to be appreciably occupied, $\sigma^2$
increases again nearly linearly till $^{146}_{64}$Gd, where a new kink appears.
From $^{146}_{64}$Gd to the proton drip line ($^{154}_{72}$Hf), the
value of $\sigma^2$ continues to increase, but now with a
smaller slope as a consequence of the higher occupancy of the $1h_{11/2}$, $2d_
{3/2}$, and $3s_{1/2}$ levels.
Indeed, along an isotonic chain the $\sigma^2$ parameter of the employed
Helm model is sensitive to the tail of the different proton single-particle wave
functions that successively contribute to the charge density.

\begin{figure}[t]
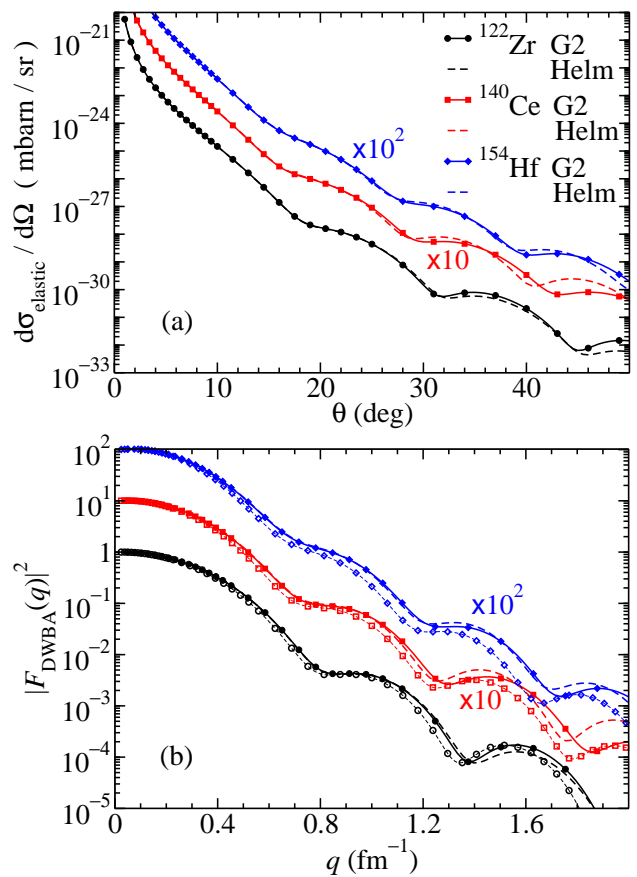

\begin{center}
\includegraphics[width=0.95\linewidth,angle=0,clip=true]{fig04a}\\
\includegraphics[width=0.95\linewidth,angle=0,clip=true]{fig04b}
\caption{\label{dcs-ff82}
(Color online) DCS for elastic electron-nucleus scattering (a) and square charge form factor (b) in $^{122}_{40}$Zr, $^{140}_{58}$Ce, and $^{154}_{72}$Hf at 500 MeV computed in DWBA. The results are shown both for the self-consistent mean-field densities of G2 (solid lines) and for the equivalent Helm distributions fitted to the G2 densities (dashed lines). In the panel (b), we also show by empty symbols the results obtained at 250 MeV using the self-consistent G2 densities.}
\end{center}
\end{figure}

We next inspect the main properties of the differential cross sections and electric charge form factors of the $N=82$ isotones. In Fig.~\ref{dcs-ff82}.a we display for three representative nuclei of the $N=82$ chain the DCS as a function of the scattering angle $\theta$. The electron beam energy is 500 MeV. The DCS is computed in the DWBA using both the self-consistent mean-field charge densities obtained with the G2 model (solid lines) and the equivalent Helm charge densities (dashed lines). The square modulus of the DWBA electric charge form factor $\vert F_{\rm DWBA}(q)\vert^2$ as a function of the momentum transfer $q=2E\sin(\theta/2)$ is shown in Fig.~\ref{dcs-ff82}.b. The empty symbols in the lower panel of this figure correspond to $\vert F_{\rm DWBA}(q) \vert^2$ computed at an electron beam energy of 250 MeV. The comparison of the results for $\vert F_{\rm DWBA}(q) \vert^2$ at $E=500$ MeV and $E=250$ MeV shows that the electric charge form factor defined in Eq.\ (\ref{fdwba}) is largely independent of the energy of the beam in the low-momentum transfer domain. Therefore, the analysis of $\vert F_{\rm DWBA}(q) \vert^2$ contains the essential trends of the elastic electron-nucleus scattering in this regime.

The dashed lines in the two panels of Fig.~\ref{dcs-ff82} correspond to the
DWBA result but using the equivalent Helm charge distributions, fitted as
explained previously, instead of the self-consistent mean-field densities. 
One can see a good agreement
at low-momentum transfers up to about 1.5 fm$^{-1}$
between the results from the original mean-field densities and from the
equivalent Helm charge densities. This fact reassures one of the ability of the
parametrized Helm distributions to describe global trends of elastic
electron-nucleus scattering at low $q$, as it was also found in Ref.\
\cite{Roc08} for isotopes.

In medium and heavy mass nuclei, the first oscillations of the DCS and of the
square charge form factor computed within the DWBA usually do not show clean
local minima but they rather show inflection points.
As we can see in Fig.~\ref{dcs-ff82}, this is the situation for the first
oscillation of the DCS and of $\vert F_{\rm DWBA}(q) \vert^2$ in the $N=82$
isotonic chain. In the absence of an explicit minimum, the first inflection
point (IP) is the best candidate to characterize the relevant properties of the
electric charge form factor at low $q$ as we discussed in Ref.\ \cite{Roc08}. 

\begin{figure}[t]
\begin{center}
\includegraphics[width=0.95\linewidth,angle=0,clip=true]{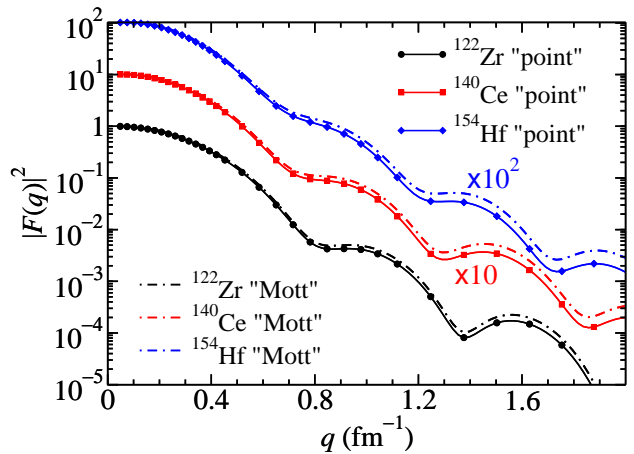}
\caption{\label{check-mott-point} (Color online) Square charge form factor in
$^{122}_{40}$Zr, $^{140}_{58}$Ce, and $^{154}_{72}$Hf at 500 MeV as predicted by
the G2 interaction calculated from Eq.~(\ref{fdwba}) (solid lines) and by using
in the denominator of Eq.~(\ref{fdwba}) the Mott DCS defined in
Eq.~(\ref{sigmamott}) instead of the DWBA point DCS (see discussion in the
text).}
\end{center}
\end{figure}

In Fig.~\ref{check-mott-point} we compare the square modulus of the
electric charge form factor as calculated in two different ways. We refer 
to the results in this figure with the label ``point'' when we show $F_{\rm
DWBA}(q)$ defined in Eq.~(\ref{fdwba}) (same solid lines shown
in Fig.~\ref{dcs-ff82}) and with the label ``Mott'' when we use in the
denominator of Eq.~(\ref{fdwba}) the Mott DCS given in Eq.~(\ref{sigmamott}).
From Fig.~\ref{check-mott-point}, one can see that $\vert F_{\rm DWBA}(q) \vert^2$ is always smaller than the result obtained using $d\sigma_{\rm Mott}/d\Omega$ in the denominator of Eq.~(\ref{fdwba}). The difference between both
results grows as the value of $q$ increases. It is also found that in this
region of low-momentum transfers (i.e., the region of main interest for our
present study), the location of the inflection
points and minima of the electric charge form factor along the $N=82$ isotonic
chain is practically independent of the choice of the denominator in
Eq.~(\ref{fdwba}).

\begin{figure}[t]
\begin{center}
\includegraphics[width=0.95\linewidth,angle=0,clip=true]{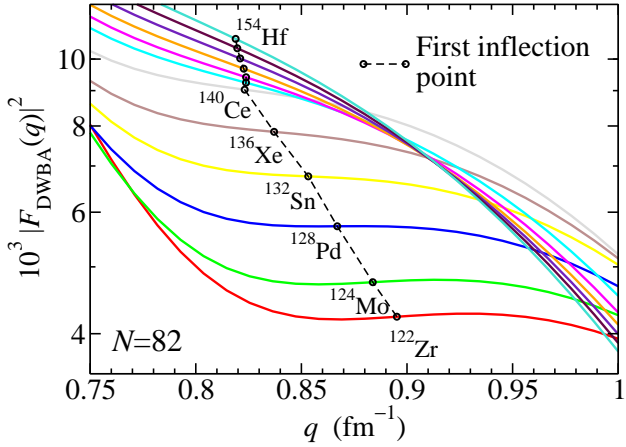}
\caption{\label{ffn82} (Color online) Evolution of the square modulus of the DWBA electric charge form factor with the momentum transfer $q$ along the $N=82$ isotonic  chain as predicted by G2 at an electron beam energy of 500 MeV. The momentum transfer corresponding to the first inflection point for each isotone is shown by circles.}
\end{center}
\end{figure}

In Fig.~\ref{ffn82} we plot $\vert F_{\rm DWBA}(q) \vert^2$ for the $N=82$ isotones
in a magnified view around the first IP. The value of $\vert F_{\rm DWBA}(q)
\vert^2$ at the first IP is depicted by circles for each nucleus. In
agreement with earlier literature \cite{Zai06}, the momentum transfer at the
first inflection point ($q_{\rm IP}$) shows an inward shifting and the value of
$\vert F_{\rm DWBA}(q_{\rm IP}) \vert^2$ shows an upward trend with increasing
mass number along the isotonic chain.

\begin{figure}[t]
\begin{center}
\includegraphics[width=0.95\linewidth,angle=0,clip=true]{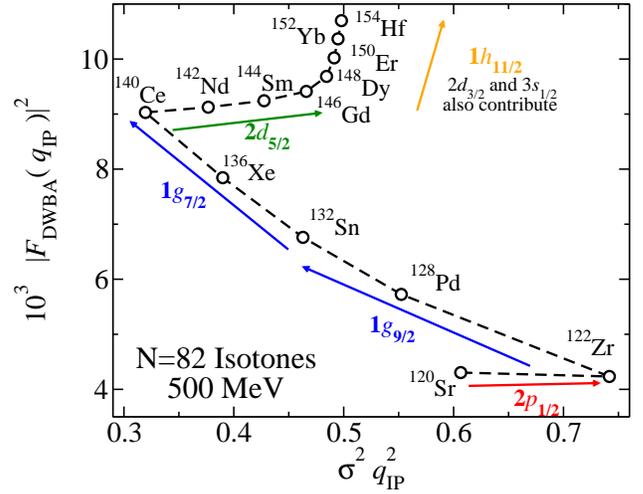}
\caption{\label{fsqn82} (Color online) Square modulus of the electric charge form factor in DWBA at the first inflection point ($q_{\rm IP}$) as a function of $\sigma^2 q_{\rm IP}^2$ predicted by the RMF model G2 for the $N=82$ isotones.}
\end{center}
\end{figure}

Let us discuss possible correlations of the DWBA charge form factor at
low-momentum transfer with the parameters $R_0$ and $\sigma$ of the equivalent
Helm charge density, as we did in our previous analysis of isotopic chains
\cite{Roc08}. If we first look at the analytical expression of the charge form
factor predicted by the Helm model, cf.\ Eq.\ (\ref{eq17}), it suggests to use
$q R_0$ and $\sigma^2q^2$ as the natural variables to investigate the variation
of this quantity. In the $q\rightarrow 0$ limit, Eq.\ (\ref{eq17}) can be
written as
\begin{equation}
F^{(H)}(q\rightarrow 0)  = 1 - \frac{1}{10}q^2\left(5\sigma^2 + R_0^2\right)+\mathcal{O}[q^4].
\label{q0helmff}
\end{equation}
This result points towards a linear correlation with the mean square radius
$\langle r_{_H}^2\rangle$ of the Helm distribution \cite{Helm56} due to the fact
that
\begin{equation}
\langle r_{_H}^2\rangle=\frac{3}{5}\left(5\sigma^2 + R_0^2\right).
\label{r2helm}
\end{equation}
However, the correlation suggested by this approximation is not fullfiled by the DWBA calculations in the relevant region of momentum transfers for our study. For this reason, we have further investigated the relation of $|F_{\rm DWBA}(q_{\rm IP})|^{2}$ with $q_{\rm IP}^2R_0^2$ and $\sigma^2q_{\rm IP}^2$ separately. We show in Fig.~\ref{fsqn82} the behavior of $|F_{\rm DWBA}(q_{\rm IP})|^{2}$ as a function of the value of $\sigma^2 q_{\rm IP}^2$ since it will be very instructive to understand the influence of the proton shell structure on elastic electron scattering in the isotonic chains.

The non-uniform variations seen in Fig.~\ref{fsqn82} along the horizontal axis
are basically due to the Helm parameter $\sigma^2$ rather than to $q^2_{\rm
IP}$. This is because of the fact that if we compare the relative change along
the isotopic chain found in the quantities $\sigma$ and $q_{\rm IP}$  (cf.\
Figs.~\ref{r0n82} and \ref{ffn82}, respectively), it is much larger in the
case of the $\sigma$ parameter. Hence, the information along the horizontal axis
of Fig.~\ref{fsqn82} is sensitive to the filling order of the single-particle
levels contributing to the charge density at the surface region.

\begin{figure}[t]
\begin{center}
\includegraphics[width=0.95\linewidth,angle=0,clip=true]{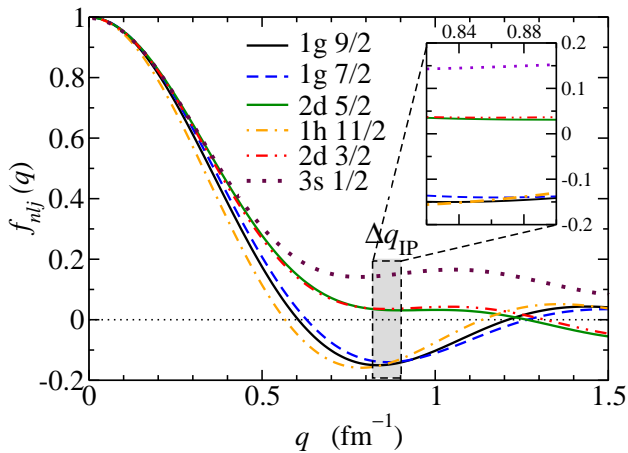}
\caption{\label{fshellsn82} (Color online) Shell contribution to the form factor of the last
occupied levels in PWBA [see Eq.\ (\ref{fsubshell})] as a function of the
momentum transfer, calculated in the nucleus $^{154}_{72}$Hf with the G2 mean field model.
The shaded region indicates the range of observed $q_{\rm IP}$ values in the calculations of the
form factor in DWBA for the $N=82$ isotonic chain.}
\end{center}
\end{figure}

The non-uniform variation shown by $|F_{\rm DWBA}(q_{\rm IP})|^{2}$ along
the vertical axis in Fig.~\ref{fsqn82} can be qualitatively understood in terms
of the single-particle contributions to the PWBA electric charge form factor. To
this end we plot in Fig.\ \ref{fshellsn82} the contribution to the PWBA form
factor from the individual proton orbitals:
\begin{equation}
f_{nlj}(q)\equiv \int d\vec{r} \mid\psi_{nlj}(\vec{r})\mid^2 e^{i \vec{q}\cdot\vec{r}} ,
\label{fsubshell}
\end{equation}
where $\psi_{nlj}(\vec{r})$ is the wave function of a proton level with quantum
numbers $n$, $l$, and $j$. Note that (\ref{fsubshell}) does not include the
occupation probability factors ($v_{nlj}$) and degeneracies ($2j+1$), i.e.,
\begin{equation}
F_{\rm PWBA}(q) = \frac{1}{Z}\sum_{nlj} (2j+1) v_{nlj} f_{nlj}(q) .
\label{ftotpwba}
\end{equation}

In Fig.\ \ref{fshellsn82} we depict $f_{nlj}(q)$ for the orbitals close to the
Fermi level in the nucleus $^{154}_{72}$Hf, which can be considered as
representative of the level scheme of the $N=82$ isotopic chain. First, one
notes that the single-particle contributions $f_{nlj}(q)$ to the PWBA electric
charge form factor do not have the same sign in the range of momentum transfers
of interest in our analysis. Therefore, strong interference effects may occur
among these single-particle contributions. In particular, we can see in Fig.\
\ref{fshellsn82} that in the region around $q_{\rm IP}$ the contributions from the 1$g_{9/2}$, 1$g_{7/2}$, and 1$h_{11/2}$ orbitals are negative, while the contributions from the 3$s_{1/2}$,
2$d_{5/2}$, and 2$d_{3/2}$ orbitals are positive. The PWBA form factor
corresponding to the underlying $Z=40$ core is negative. Therefore, when the
$1g_{9/2}$ and $1g_{7/2}$ orbitals are occupied---in passing from
${}_{40}^{122}$Zr to ${}_{58}^{140}$Ce---the square modulus of the PWBA form
factor increases.
When on top of this configuration, the $2d_{5/2}$ orbital is filled in
${}_{64}^{146}$Gd, the square modulus of the PWBA charge form factor decreases
due to the positive sign of the contribution of this level around $q_{\rm IP}$. This simple pattern
in the uniform filling picture is slighlty modified due to the pairing
correlations that introduce additional mixing with the contributions from the
$2d_{3/2}$, $3s_{1/2}$, and $1h_{11/2}$ orbitals. In spite of this, the simple
PWBA description is quite useful to help us interpret the changes of $|F_{\rm
DWBA}(q_{\rm IP})|^{2}$ from ${}_{40}^{122}$Zr to ${}_{64}^{146}$Gd. The
subsequent increase shown by $|F_{\rm DWBA}(q_{\rm IP})|^{2}$ from 
${}_{64}^{146}$Gd to ${}_{72}^{154}$Hf can also be understood in this schematic
picture since the PWBA charge form factor of ${}_{64}^{146}$Gd is globally
negative and the additional contribution of the $1h_{11/2}$ orbital also is
negative for $q$ values near $q_{\rm IP}$.

The discussed theoretical results pinpoint the importance of the
filling order of the proton single-particle levels in elastic electron
scattering off exotic nuclei. This fact suggests that future
experiments such as those planned in the upgrades of the GSI and RIKEN
facilities may become excellent probes of the shell structure of
exotic nuclei. However, the small values of the cross sections and the short half-lives and small production rates of many nuclei along an isotonic chain can be strong limitations for such kind of measurements in practice.

Regarding the relation of $|F_{\rm DWBA}(q_{\rm IP})|^{2}$ with
$q_{\rm IP}^2R_0^2$, we have not found a simple behavior. In spite of
this, we have observed that $|F_{\rm DWBA}(q_{\rm IP})|^{2}$ and the
square of the Helm radius $R_0^2$ show a rather similar behavior as a
function of the mass number in the isotonic chain.
\begin{figure}[t]
\begin{center}
\includegraphics[width=0.95\linewidth,angle=0,clip=true]{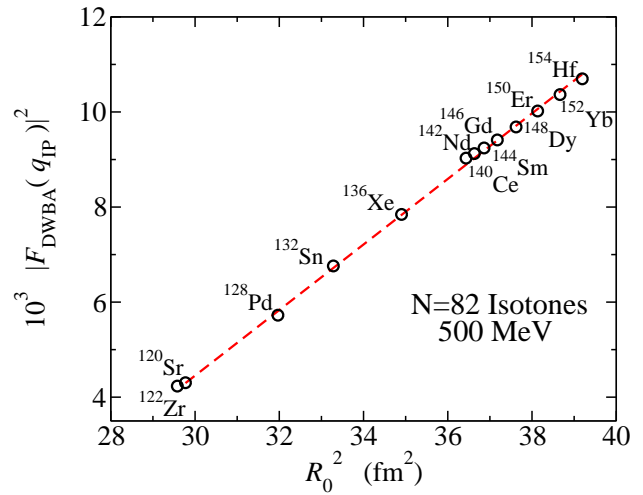}
\caption{\label{fvsR0n82} (Color online) Square modulus of the electric charge form factor in DWBA at the first inflection point ($q_{\rm IP}$) as a function of $R_0^2$ predicted by the RMF model G2 for the $N=82$ isotones.}
\end{center}
\end{figure}
This suggests plotting $|F_{\rm DWBA}(q_{\rm IP})|^{2}$ against $R_0^2$,
which we do in Fig. \ref{fvsR0n82}. One observes a good linear correlation
between both quantities. This correlation indicates that the parameter of the
Helm model which measures the size of the bulk part of the density profile of
each isotone governs the magnitude of the electric charge form factor at low momentum transfer.

We have found that fitting the calculated PWBA electric charge form factor with an extended Helm model \cite{Frie86} instead of the simple Helm model of Section II.A, leads to similar conclusions on the behavior of $R_0$ and $\sigma^2$ along the isotonic chain. It is also worth noticing that the influence of the proton shell structure on $|F_{\rm DWBA}(q_{\rm IP})|^{2}$ (i.e., the changes shown by $|F_{\rm DWBA}(q_{\rm IP})|^{2}$ along the vertical axis of Fig.~\ref{fsqn82} as the different proton orbitals are being filled) is independent of the Helm model used to fit the PWBA electric charge form factor.
To conclude this section, we would like to note that some of the details of the
predicted single-particle energies, energy gaps, and filling order of the
orbitals change to some extent if in our calculations we use other RMF models or
Skyrme forces instead of the G2 interaction. In particular, this is due to the
fact that we are exploring regions of the nuclear chart beyond the
region where the parameters of these effective nuclear interactions have been
calibrated. However, the basic conclusion to be
emphasized, i.e., the manifest sensitivity of some electron scattering
observables to the proton shell structure of the isotones, is a robust feature
that comes out regardless of the effective nuclear interaction.

\subsection{$N=50$ isotonic chain}

\begin{figure}[t]
\begin{center}
\includegraphics[width=0.95\linewidth,angle=0,clip=true]{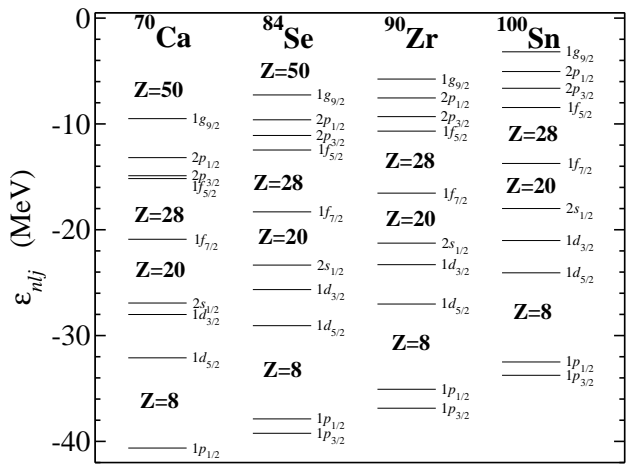}
\caption{\label{spln50} Energy of the proton  single-particle levels for $^{70}_{20}$Ca, $^{84}_{34}$Se, $^{90}_{40}$Zr, and $^{100}_{50}$Sn as computed with the G2 parameter set.}
\end{center}
\end{figure}

The more relevant proton single-particle orbitals for our study of the $N=50$ chain are the $1f_{7/2}$, $1f_{5/2}$, $2p_{3/2}$, $2p_{1/2}$, and $1g_{9/2}$ orbitals. They cover two major shells between Ca and Sn. This set of proton energy levels computed with the G2 parametrization is displayed in Fig.\ \ref{spln50} for some selected isotones. We can see that these levels move up in energy, roughly as a whole, when the mass number increases in going from proton-deficient nuclei ($^{70}_{20}$Ca) to stable nuclei ($^{84}_{34}$Se, $^{90}_{40}$Zr) and to proton drip-line nuclei ($^{100}_{50}$Sn).

\begin{figure}[t]
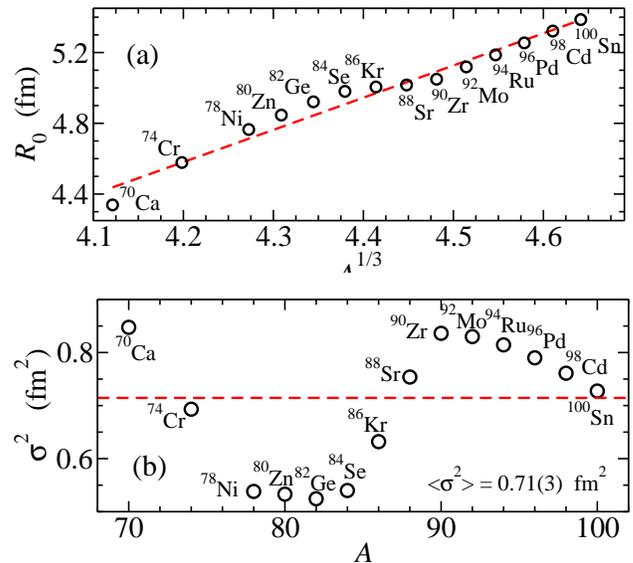

\begin{center}
\includegraphics[width=0.95\linewidth,angle=0,clip=true]{fig11a}
\vspace*{3mm}

\includegraphics[width=0.95\linewidth,angle=0,clip=true]{fig11b}
\caption{\label{r0n50}
(Color online) Mass-number dependence of the Helm parameter $R_{0}$ predicted by the covariant mean field model G2 in the $N=50$ isotonic  chain (a). Mass-number dependence of the Helm parameter $\sigma^2$ (b). The average value is depicted by a horizontal dashed line.}
\end{center}
\end{figure}

The parameters $R_0$ and $\sigma^2$ of the Helm model distributions fitted to the mean-field charge densities of the $N=50$ isotones are displayed in Fig.~\ref{r0n50} and given in Table \ref{r-sp-table}. The global features are similar to the case of the $N=82$ chain. The $R_0$ parameter, which represents the effective location of the surface of the nucleus, approximately follows a linear trend with $A^{1/3}$. The mass-number dependence of $\sigma^2$ again displays a non-uniform trend, originated by the filling of the different proton single-particle orbitals.
We see that $\sigma^2$ decreases in filling the $1f_{7/2}$ shell from
$^{70}_{20}$Ca to $^{78}_{28}$Ni, it remains roughly constant when the
$1f_{5/2}$ level is being filled up to $^{84}_{34}$Se, it increases when the
$2p_{3/2}$ and $2p_ {1/2}$ shells are occupied till $^{90}_{40}$Zr, and it then
decreases until the proton drip-line nucleus $^{100}_{50}$Sn is reached by
filling the $1g_ {9/2}$ level. Therefore, the more abrupt (smaller~$\sigma$) equivalent charge densities predicted by the G2 model in the $N=50$ isotonic chain correspond to nuclei between the doubly-magic, proton-deficient
$^{78}_{28}$Ni nucleus and the more stable $^{84}_{34}$Se nucleus, where mainly
the $1f_{5/2}$ shell has been filled. In these nuclei, the occupancy of the
$2p_{3/2}$, $2p_ {1/2}$, and $1g_ {9/2}$ levels due to the pairing correlations
is rather small.

\begin{figure}[b]
\begin{center}
\includegraphics[width=0.95\linewidth,angle=0,clip=true]{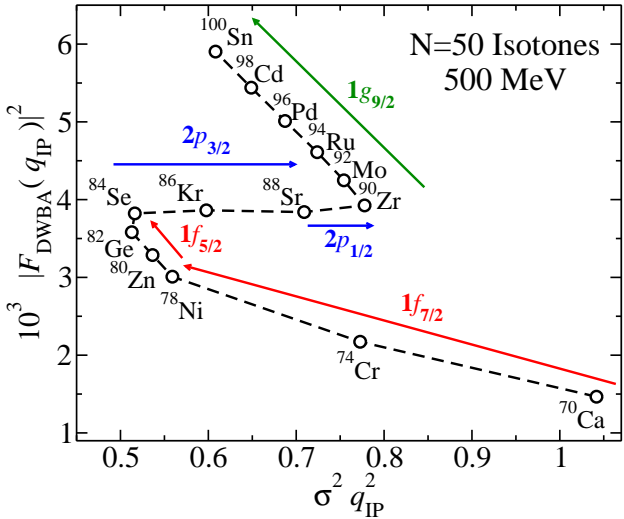}
\caption{\label{fsqn50} (Color online) Square modulus of the electric charge form factor in  DWBA at the first inflection point ($q_{\rm IP}$) as a function of $\sigma^2 q_{\rm IP}^2$ predicted by  the RMF model G2 for the $N=50$ isotones.}
\end{center}
\end{figure}

The square modulus of the DWBA electric charge form factor for an electron beam
energy of 500 MeV is displayed against $\sigma^2q^2_{\rm IP}$ in Fig.~\ref{fsqn50}.
As in the case of the $N=82$ isotones, the behavior of $\sigma^2 q^2_{\rm IP}$ is
dominated by the Helm parameter $\sigma$. This is because the relative variation
of $\sigma^2$ (see Fig.~\ref{r0n50}) is much larger than the relative variation
of $q^2_{\rm IP}$ along the isotonic chain. The change of $\vert F_{\rm DWBA}(q_{\rm IP}) \vert^2$ along the $N=50$ chain shows, globally, an increasing trend with the mass number.
We can appreciate in Fig.~\ref{fsqn50} that although the variation of $\vert F_{\rm DWBA}(q_{\rm IP}) \vert^2$ is almost linear when a specific proton orbital is being occupied, drastic changes of slope take place when a new shell starts to be significantly occupied. Recalling the simplified PWBA picture, cf.\ Eqs.\ (\ref{fsubshell}) and (\ref{ftotpwba}), we find that in the region of $q$ values around $q_{\rm IP}$ the contribution to the electric charge form factor from the $1f$ and $1g$ orbitals is negative, while the contribution from the $2p$ orbitals is positive. This fact is consistent with the behavior shown by $\vert F_{\rm DWBA}(q_{\rm IP}) \vert^2$ in Fig.~\ref{fsqn50}. That is, $\vert F_{\rm DWBA}(q_{\rm IP}) \vert^2$ increases in passing from $^{70}_{20}$Ca to $^{84}_{34}$Se, basically due to the filling of the $1f_{7/2}$ and $1f_{5/2}$ shells,
\begin{figure}[t]
\begin{center}
\includegraphics[width=0.95\linewidth,angle=0,clip=true]{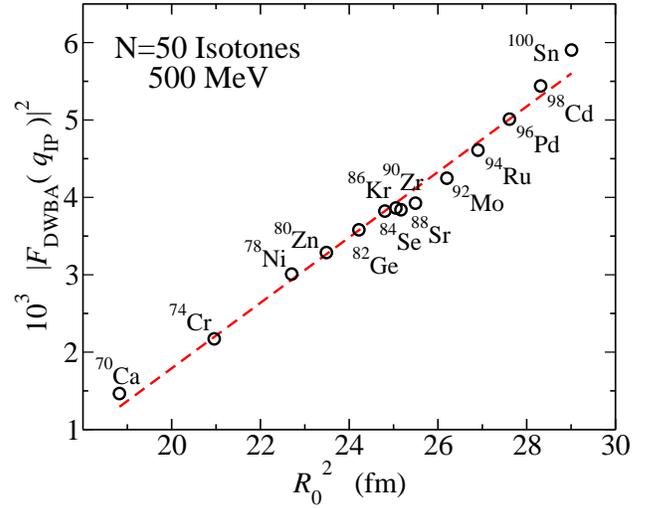}
\caption{\label{fvsR0n50} (Color online) Square modulus of the electric charge form factor in  DWBA at the first inflection point ($q_{\rm IP}$) as a function of $R_0^2$ predicted by  the RMF model G2 for the $N=50$ isotones.}
\end{center}
\end{figure}
and then its value is practically quenched up~to $^{90}_{40}$Zr because the
$2p_{3/2}$ and $2p_{1/2}$ orbitals contribute with opposite sign to the $1f$ orbitals. When the
$1g_{9/2}$ level is appreciably occupied in approaching the proton drip line,
the value of $\vert F_{\rm DWBA}(q_{\rm IP}) \vert^2$ increases again with a
nearly constant rate. Finally, in Fig.~\ref{fvsR0n50} we see that $\vert F_{\rm
DWBA}(q_{\rm IP}) \vert^2$ of the $N=50$ isotones shows a good linear
correlation with the square of the Helm parameter~$R_0$.

\subsection{$N=14$ isotonic chain}

In this section we discuss the lightest isotonic chain analyzed in our work.
Although the present findings are to be taken with some reservations because
the mean-field approach is not best suited for light-mass exotic nuclei, we note
that similar general trends to those observed
\begin{figure}[t]
\begin{center}
\includegraphics[width=0.95\linewidth,angle=0,clip=true]{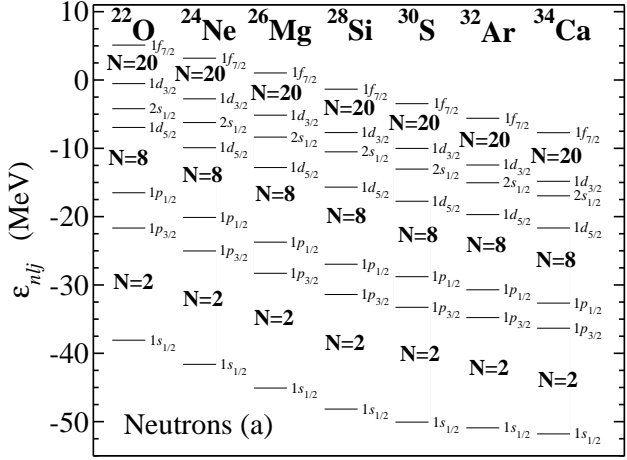}\\
\vspace{3mm}
\includegraphics[width=0.95\linewidth,angle=0,clip=true]{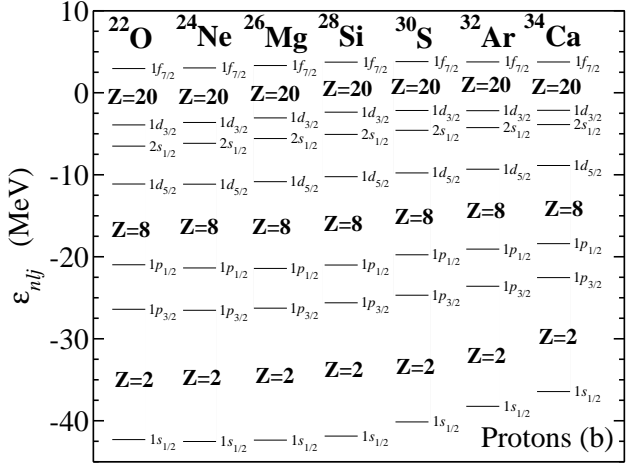}
\caption{\label{spln14} Energy of the neutron (a) and proton (b) single-particle levels for $^{22}_{8}$O, $^{24}_{10}$Ne, $^{26}_{12}$Mg, $^{28}_{14}$Si, $^{30}_{16}$S, $^{32}_{18}$Ar and $^{34}_{20}$Ca as computed with the G2 parameter set.}
\end{center}
\end{figure}
in the heavier-mass isotonic chains also appear in the $N=14$ chain.

In Fig.~\ref{spln14} we display the neutron [panel (a)] and proton [panel (b)] 
single-particle levels computed with the G2 interaction for the $N=14$
isotonic chain, from the very proton-deficient nucleus $^{22}_{8}$O to the very
proton-rich nucleus $^{34}_{20}$Ca. As expected, the neutron single-particle
levels become more bound with increasing mass number. In addition to the
prominent energy gap at $N=8$ seen in the whole chain, it may be noticed that the
$1d_{5/2}$ neutron level becomes progressively more isolated when the mass
number increases, which points to some magic character of the neutron number
$N=14$ in the calculation with the G2 model. This magic character is confirmed
by the vanishing neutron pairing gap found in our calculation from $^{26}_{12}$Mg to $^{34}_{20}$Ca.
It may be observed that the G2 model also predicts a slightly magic character of
the neutron number $N=16$ towards the neutron drip line.
Actually, we see in Fig.~\ref{spln14} that the relatively magic trend of
$N=14$ increases from the proton-poor side ($^{22}_{8}$O) to the proton-rich
side ($^{34}_{20}$Ca) of the chain, while the somewhat magic trend of
$N=16$ decreases from $^{22}_{8}$O to $^{34}_{20}$Ca.

The more relevant proton single-particle orbitals for our study of the $N=14$
chain belong to the $s$-$d$ major shell. The energy levels of this proton major
shell (see Fig.~\ref{spln14}.b) lie approximately at the same
energy for all the nuclei from $^{22}_{8}$O to $^{34}_{20}$Ca, with roughly
constant energy gaps. The $1d_{5/2}$ and $2s_{1/2}$ proton levels exhibit
a considerable energy gap between them in this isotonic chain according to the predictions
of the G2 model. It is to be mentioned that due to the
pairing correlations, the proton levels $1f_ {7/2}$ and $1f_{5/2}$ (the latter
is not displayed in Fig.~\ref{spln14}) also play some role in our calculation of
the mean-field charge densities. These levels simulate to a certain extent the
effect of the continuum due to their quasi-bound character owing to the Coulomb and
centrifugal barriers \cite{Est01a}.

\begin{figure}[t]
\begin{center}
\includegraphics[width=0.95\linewidth,angle=0,clip=true]{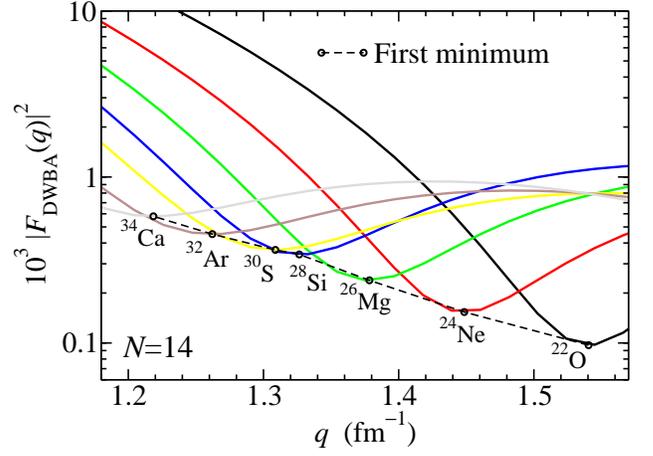}
\caption{\label{ffn14} (Color online) Evolution of the square modulus of the DWBA electric charge form factor with the momentum transfer $q$ along the $N=14$ isotonic chain as predicted by G2 at an electron beam energy of 500 MeV. The momentum transfer corresponding to the first minimum for each isotone is shown by circles.}
\end{center}
\end{figure}

\begin{figure}[t]
\begin{center}
\includegraphics[width=0.95\linewidth,angle=0,clip=true]{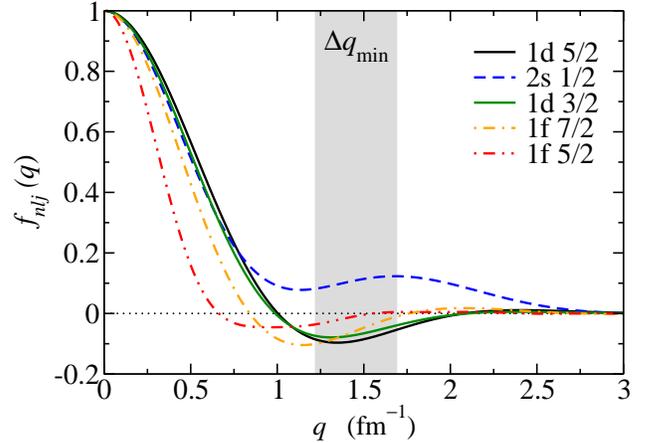}
\caption{\label{fshellsn14} (Color online) Shell contribution to the form factor of the last occupied levels in PWBA [see Eq.\ (\ref{fsubshell})] as a function of the momentum transfer. The shaded region indicates the range of observed $q_{\rm min}$ values in the calculations of the form factor in DWBA for the $N=14$
isotonic chain.}
\end{center}
\end{figure}

In Fig.~\ref{ffn14} we display a magnified view of $\vert F_{\rm DWBA}(q)\vert^2$ against the momentum transfer for the $N=14$ isotonic chain.
We see that in this chain of lower mass, the inflection point that was found after the first oscillation of the charge form factor in the heavier chains $N=50$ and $N=82$ becomes a clearly well defined local minimum. Thus, for the discussion of the $N=14$ chain we focus on the properties of $\vert F_{\rm DWBA}(q)\vert^2$ at its first minimum ($q_{\rm min}$). The value of $\vert F_{\rm DWBA}(q_{\rm min}) \vert^2$ (shown by the circles in Fig.~\ref{ffn14}) increases
when the value of the momentum transfer at the first minimum decreases, i.e., $\vert F_{\rm DWBA}(q_{\rm min}) \vert^2$  grows with increasing mass number in the chain. Though the increase of $\vert F_{\rm DWBA}(q_{\rm min}) \vert^2$ in Fig.~\ref{ffn14} is roughly linear with $q_{\rm min}$, one notes a kink at the point corresponding to the $^{28}_{14}$Si nucleus. As we can realize from Fig.~\ref{fshellsn14} in the schematic PWBA picture, this kink is originated by cancellation effects between the opposite contributions to the charge form factor around $q_{\rm min}$ coming from the single-particle wave function of the $1d$ proton level (negative contribution) and of the $2s$ proton level (positive contribution).
\begin{figure}[b]
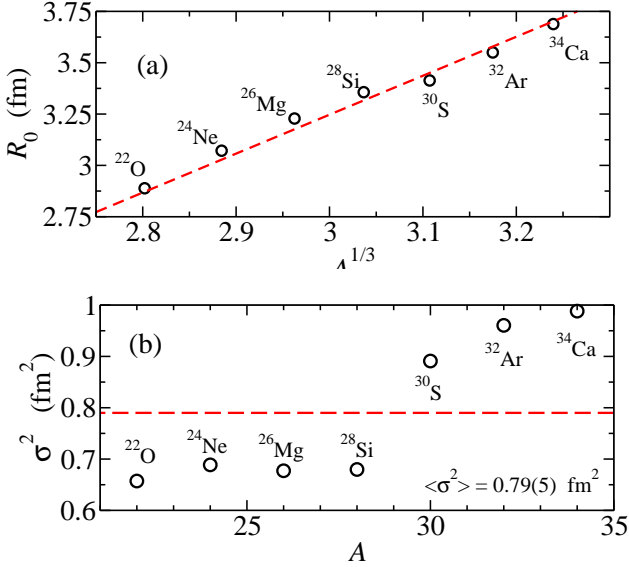

\begin{center}
\includegraphics[width=0.95\linewidth,angle=0,clip=true]{fig17a}
\vspace*{3mm}

\includegraphics[width=0.95\linewidth,angle=0,clip=true]{fig17b}
\caption{\label{r0n14} (Color online) Mass-number dependence of the Helm parameter $R_{0}$ predicted by the covariant mean field model G2 in the $N=50$ isotonic  chain (a). Mass-number dependence of the Helm parameter $\sigma^2$ (b). The average value is depicted by a horizontal dashed line.}
\end{center}
\end{figure}

The Fig.~\ref{r0n14}.a shows that the parameter $R_0$ of the equivalent Helm
charge densities displays, as in the heavier isotonic
chains, an overall linear increasing trend with $A^{1/3}$. In
turn, the variation of the Helm parameter $\sigma$ reflects the underlying shell
structure of the mean-field charge densities. In the 
Fig.~\ref{r0n14}.b, we see that the value of $\sigma^2$ remains
almost constant between $^{22}_{8}$O and $^{28}_{14}$Si when mainly the
$1d_{5/2}$ shell is being filled. From $^{30}_{16}$S on, the $2s_{1/2}$ and
$1d_{3/2}$ levels start to be appreciably occupied and $\sigma^2$ starts
increasing almost linearly with $A$ till the proton-drip line nucleus
$^{34}_{20}$Ca. The numerical value of both Helm model parameters for the $N=14$ 
isotonic chain can be found in Table \ref{r-sp-table}. 

\begin{figure}[t]
\vspace{-10mm}
\begin{center}
\includegraphics[width=0.80\linewidth,clip=true,angle=-90]{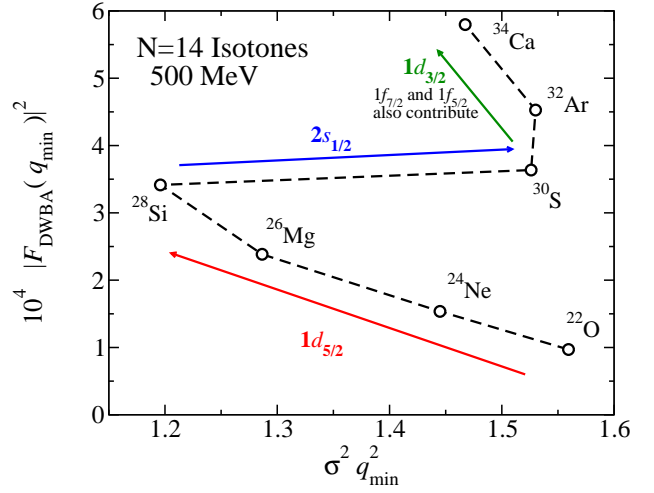}
\caption{\label{fsqn14} (Color online) Square modulus of the electric charge form factor in DWBA at the first minimum ($q_{\rm min}$) as a function of $\sigma^2 q_{\rm min}^2$ predicted by the RMF model G2 for the $N=14$ isotones.}
\end{center}
\end{figure}

\begin{figure}[b]
\begin{center}
\includegraphics[height=0.75\linewidth,angle=0,clip=true]{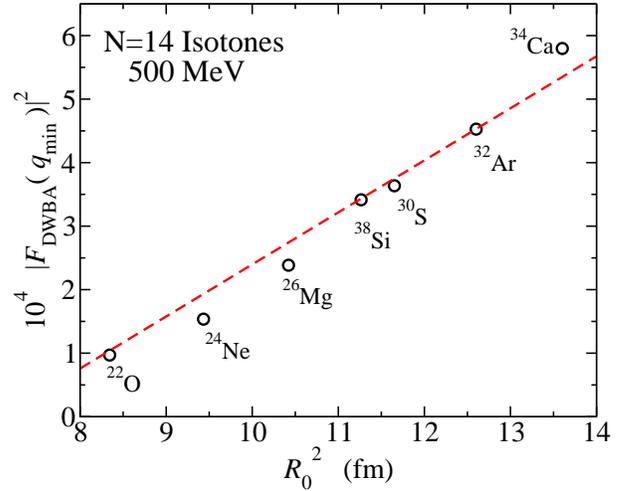}
\caption{\label{fvsR0n14} (Color online) Square modulus of the electric charge form factor in  DWBA at the first minimum ($q_{\rm min}$) as a function of $R_0^2$ predicted by  the RMF model G2 for the $N=14$ isotones.}
\end{center}
\end{figure}

The influence of the discussed proton shell structure on the electric charge
form factor at the first minimum for the $N=14$ isotones is obvious in Fig.~\ref{fsqn14}, which
displays $\vert F_{\rm DWBA}(q_{\rm min}) \vert^2$ against the value of
$\sigma^2 q_{\rm min}^2$. Finally, if we analyze the variation of
$\vert F_{\rm DWBA}(q_{\rm min}) \vert^2$ with the Helm parameter $R_0^2$,
a correlation is found between both quantities (cf.\ Fig.~\ref{fvsR0n14}),
though now this correlation is less linear than in the heavier isotonic
chains $N=82$ and $N=50$ (cf.\ Figs.~\ref{fvsR0n82} and \ref{fvsR0n50}).


\section{Summary and conclusions}

In this work we have explored some of the information about nuclear structure
that can be obtained from elastic electron scattering in isotones.
Due to the fact that the number of protons changes along an isotonic chain, this
study primarily probes the effect of the different proton single-particle shells
on the elastic electron scattering observables.

We have computed the DWBA differential cross section at an electron beam energy
of 500 MeV using the charge densities calculated self-consistently with
the covariant nuclear mean-field model G2 \cite{G2,Ser97}. The electric charge
form factor has been obtained by taking the ratio of the DWBA differential cross
section with the DWBA point nucleus differential cross section. The so-defined
electric charge form factor is practically independent of the electron beam
energy in the low-momentum transfer regime.

We have paid special attention to the electric charge form
factor taken at the momentum transfer $q_{\rm IP}$ of the first
inflection point, for the $N=50$ and $N=82$ chains, or at the momentum
transfer $q_{\rm min}$ of the first
minimum for the $N=14$ chain. In agreement with earlier
literature, we have found that the values of $q_{\rm IP}$ ($q_{\rm min}$)
shift inwards (i.e., they become smaller) and that the values of
$\vert F_{\rm DWBA}(q_{\rm IP}) \vert^2$ ($\vert F_{\rm DWBA}(q_{\rm min})
\vert^2$) increase when the atomic number increases from the neutron
drip line to the proton drip line of the isotonic chain.

The results reveal that along an isotonic chain the DWBA electric charge form
factor of each nucleus is extremely sensitive to the underlying proton shell
structure. In the simpler PWBA picture, we have found that a particular proton
shell may contribute to the electric charge form factor with positive or
negative sign in the momentum transfer region of interest in our analysis. The
contributions from levels with more than one radial node and small orbital
angular momenta have opposite sign to the contributions from levels without
radial nodes and large angular momenta. As a consequence, cancellation
effects can appear when the different proton shells are successively occupied.
Although this description is to some extent masked by pairing correlations, a
similar situation is found in the DWBA calculations of the electric charge form
factor. Therefore, the rate of change of $\vert F_{\rm DWBA}(q_{\rm IP})
\vert^2$ or $\vert F_{\rm DWBA}(q_{\rm min}) \vert^2$ with the proton number along
an isotonic chain may vary substantially when a new single-particle orbital
enters the nucleus. This suggests that scattering experiments performed on
isotones can be effective probes of the proton nuclear shell structure
of stable and unstable nuclides.

To investigate the dependence of the first inflection point or first minimum
of the electric charge form factor with the basic properties of the
charge distribution, we have parametrized the mean-field densities by the
Helm model. We have found that the Helm parameter $R_0$, which measures the mean
position of the surface of the charge density, increases with the mass
number following roughly an $A^{1/3}$ law. The Helm parameter $\sigma$,
which is related with the surface thickness of the charge distribution, encodes
information from the underlying proton shell structure and, as a consequence, it
does not show a definite trend with the mass number. We have also found that
at low-momentum tranfers the square modulus of the DWBA electric charge form
factor is accurately reproduced if the mean-field densities are replaced by the
fitted Helm densities.

In a previous work \cite{Roc08}, we noted that correlations between the DWBA
electric charge form factor at the first inflection point (or minimum) and the
parameters of the Helm densities can provide information about global
features of electron scattering along isotopic chains. In the isotonic chains
analyzed in the present work, we find that $\vert F_{\rm DWBA}(q_{\rm IP})
\vert^2$ ($\vert F_{\rm DWBA}(q_{\rm min}) \vert^2$) shows a rather good linear
correlation with the Helm parameter $R_0^2$, specially in the heavier isotonic
chains, while there is no regular behavior with the Helm parameter $\sigma^2$
because of its dependence on the last occupied proton orbitals. It should be
pointed out that shell effects encoded in the DWBA electric charge form factor
are magnified if $\vert F_{\rm DWBA}(q_{\rm IP}) \vert^2$ ($\vert F_{\rm
DWBA}(q_{\rm min}) \vert^2$) is plotted against $\sigma^2 q_{\rm IP}^2$
($\sigma^2 q_{\rm min}^2$) along an isotonic chain, providing interesting
insights into the proton shell structure of the nuclei of the chain.

In summary, the found theoretical results indicate that electron
scattering in isotonic chains can be a useful tool to probe the single-particle
shell structure of exotic nuclei and, in particular, to provide some insight
about the filling order and occupancy of the different valence proton orbitals.
Experimentally, the investigation is more difficult due to the limitations
arising from small production rates, short half-lives, and small cross
sections when one deals with unstable nuclei
\cite{sud11,Kat03,GSI02,Sim04,Sud05,Wak08,Sud09,Sim07,Ant11}.

\acknowledgments
M.C. and X.V. acknowledge the support of the Consolider Ingenio 2010
Programme CPAN CSD2007-00042, Grant No. FIS2011-24154 from MICINN and
FEDER, and Grant No. 2009SGR-1289 from Generalitat de Catalunya.
F.S. acknowledges support from the Spanish Ministerio de Ciencia e Innovaci\'on
and FEDER (project No.\ FPA2009-14091-C02-01) and from the Generalitat de Catalunya
(grant SGR 2009-276).
X.R. acknowledges support of the Italian Research Project ``Many-body
theory of nuclear systems and implications on the physics of neutron
stars'' (PRIN 2008).


%

\begin{thebibliography}{99}
%
%
\bibitem{Hof56}  R.~Hofstadter, 
                 Rev.\ Mod.\ Phys.\ {\bf 28}, 214 (1956).

\bibitem{Don75}  T.~W.~Donnelly and 
                 J.~D.~Walecka, 
                 Annu.\ Rev.\ Nuc.\ Part.\ Sci.\ {\bf 25}, 329 (1975). 

\bibitem{Don84}  T.~W.~Donnelly and 
                 I.~Sick, 
                 Rev.\ Mod.\ Phys.\ {\bf 56}, 461 (1984).

\bibitem{Moya86} E.~Moya de Guerra, 
                 Phys.\ Rep.\ {\bf 138}, 293 (1986).

\bibitem{Sick01} I.~Sick, 
                 Prog.\ Part.\ Nucl.\ Phys.\ {\bf 47}, 245 (2001).

\bibitem{Vri87}  H.~de Vries, 
                 C.~W.~de Jager, and 
                 C.~de Vries, 
                 At.\ Data Nucl.\ Data Tables, {\bf 36}, 495 (1987).

\bibitem{Fri95}  G.~Fricke, 
                 C.~Bernhardt, 
                 K.~Heiling, 
                 L.~A.~Schaller, 
                 L.~Shellenberg, 
                 E.~B.~Shera, and 
                 C.~W.~de Jager, 
                 At.\ Data  Nucl.\ Data Tables  {\bf 60}, 177 (1995).

\bibitem{Ang04}  I.~Angeli, 
                 At.\ Data Nucl.\ Data Tables {\bf 87}, 185 (2004).

\bibitem{ENAM08} {\it Topical Issue on the Fifth International Conference on 
                 Exotic Nuclei and Atomic Masses``ENAM 08''},
                 edited by J.~\"Ayst\"o, W.~Nazarewicz,
                 M.~Pf\"utzner, and C.~Signorini,                    
                 Eur.\ Phys.\ J.\ {\bf A 42}, No.~3 (2009).

\bibitem{Tan95}  I.~Tanhihata, Prog.\ Part.\ Nucl.\ Phys.\ {\bf 35}, 505 (1995).
 
\bibitem{Gei95}  H.~Geissel, 
                 G.~M\"uzenberg, and 
                 R.~Riisager, 
                 Annu.\ Rev.\ Nucl.\ Part.\ Sci.\ {\bf 45}, 163 (1995).
 
\bibitem{Mue01}  A.~Mueller, 
                 Prog.\ Part.\ Nucl.\ Phys.\ {\bf 46}, 359 (2001).

\bibitem{xia02}  J.~W.~Xia {\it et al.}, 
                 Nucl.\ Inst.\ and Meth.\ in Phys.\ Res.\ {\bf A 488} 11 (2002). 
  

\bibitem{xia07}  J.~W.~Xia {\it et al.}, 
                 Construction and Commisioning of HIRFL-CSR, ``APAC2007'',
Indore, India, January 2007.

\bibitem{sud11}  T.~Suda, 
                 J.\ Phys.: Conf.\ Ser.\ {\bf 267} 012008 (2011).

\bibitem{Kat03}  K.~Katayama, 
                 T.~Suda, and 
                 I.~Tanihata, 
                 Phys.\ Scr.\ {\bf T104}, 129 (2003).

\bibitem{GSI02}  An International Accelerator Facility for Beams of Ions and Antiprotons, GSI report 2006. 
                 http://www.gsi.de/GSI-Future/cdr/

\bibitem{Sim04}  H.~Simon in 
                 {\it Proceedings of the International Workshop XXXII on Gross Properties of Nuclei and Nuclear Excitations}, 
                 edited by M.~Buballa, J.~Knoll, W.~N\"orenberg, B.-J.~Schaefer, and J.~Wambach, GSI, Darmstad, (2004), p.290.


\bibitem{Sud05}  T.~Suda and
                 M.~Wakasugi, Prog.\ Nucl.\ Phys.\ {\bf 55}, 417 (2005).

\bibitem{Wak08}  M.~Wakasugi,
                 T.~Emoto,
                 Y.~Furukawa,
                 K.~Ishii,
                 S.~Ito,
                 T.~Koseki,
                 K.~Kurita,
                 A.~Kuwajima,
                 T.~Masuda,
                 A.~Morikawa,
                 M.~Nakamura,
                 A.~Noda,
                 T.~Ohnishi,
                 T.~Shirai,
                 T.~Suda,
                 H.~Takeda,
                 T.~Tamae,
                 H.~Tongu,
                 S.~Wang, and
                 Y.~Yano,
                 Phys.\ Rev.\ Lett.\ {\bf 100}, 164801 (2008).

\bibitem{Sud09} T. Suda, M. Wakasugi, T. Emoto, K. Ishii, S. Ito, K. Kurita, A. Kuwajima, A. Noda, T. Shirai, T. Tamae, H. Tongu, S. Wang, and Y. Yano,
                 Phys.\ Rev.\ Lett.\ {\bf 102}, 102501 (2009).

\bibitem{Sim07}  H.~Simon, 
                 Nucl.\ Phys.\ {\bf A787}, 102 (2007).

\bibitem{Ant11}  A.~N.~Antonov et al,
                 Nucl.\ Instr.\ and Meth.\ {\bf A637}, 60 (2011).


\bibitem{Gar99}  E.~Garrido and E.~Moya de Guerra,
                 Nucl.\ Phys.\ {\bf A650}, 387 (1999);
                 Phys.\ Lett.\ {\bf B488}, 68 (2000).

\bibitem{Ant05}  A.~N.~Antonov,
                 D.~N.~Kadrev,
                 M.~K.~Gaidarov,
                 E.~Moya de Guerra,
                 P.~Sarriguren,
                 J.~M.~Udias, 
                 V.~K.~Lukyanov, 
                 E.~V.~Zemlyanaya, and 
                 G.~Z.~Krumova, 
                 Phys.\ Rev.\ {\bf C72}, 044307 (2005).

\bibitem{Sar07}  P.~Sarriguren, 
                 M.~K.~Gaidarov, 
                 E.~Moya de Guerra, and
                 A.~N.~Antonov, 
                 Phys.\ Rev.\ {\bf C76}, 044322 (2007). 

\bibitem{Karat07} S.~Karataglidis and                  K.~Amos,  
                 Phys.\ Lett.\ {\bf B650}, 148 (2007).

\bibitem{Bertu07} C.~A.~Bertulani, 
                 J.~Phys.\ {\bf G34}, 315 (2007). 

\bibitem{Zai04}  Z. Wang and
                 Z. Ren,
                 Phys.\ Rev.\ {\bf C70}, 034303 (2004);
                 Phys.\ Rev.\ {\bf C71}, 054323 (2005).

\bibitem{Roc08}  X.~Roca-Maza,
                 M.~Centelles, 
                 F.~Salvat, and 
                 X.~Vi\~nas,
                 Phys.\ Rev.\ {\bf C78}, 044332 (2008). 

\bibitem{Zai06}  Z. Wang
                 Z. Ren, and
		 Y. Fan,
                 Phys.\ Rev.\ {\bf C73}, 014610 (2006).

\bibitem{Amos04} K.~Amos, 
                 S.~Karataglidis and 
                 J.~Dobaczewski,
                 Phys.\ Rev.\ {\bf C70}, 024607 (2004).

\bibitem{Sud04}  T.~Suda in 
                 {\it Proceedings of the International Workshop XXXII on Gross Properties of Nuclei and Nuclear Excitations}, 
                 edited by M.~Buballa, J.~Knoll, W.~N\"orenberg, B.-J.~Schaefer, and J.~Wambach, GSI, Darmstad, (2004), p.235.

\bibitem{Yen54}  D.~R.~Yennie, 
                 D.~G.~Ravenhall, and 
                 R.~N.~Wilson, 
                 Phys.\ Rev.\  {\bf 95}, 500 (1954).

\bibitem{Heisenberg81}  J.~H.~Heisenberg,
                 Adv.\ Nucl.\ Phys.\ {\bf 12}, 61 (1981).

\bibitem{Nishimura85} M.~Nishimura, E.~Moya de Guerra, and D.~W.~L.~Sprung,
                 Nucl.\ Phys.\ {\bf A435}, 523 (1985).

\bibitem{Udias93} J.~M. Ud{\'\i}as, P. Sarriguren, E. Moya de Guerra, 
                 E. Garrido, and J.~A. Caballero,
                 Phys.\ Rev.\ {\bf C48}, 2731 (1993).


\bibitem{Caballero98} J.~A. Caballero, T.~W. Donnelly, E. Moya de Guerra, 
                 and J.~M. Ud{\'\i}as,
                 Nucl.\ Phys.\ {\bf A643}, 189 (1998).

\bibitem{Chu10}  Y. Chu,
                 Z. Ren,
                 Z. Wang, and
                 T. Dong,
                 Phys.\ Rev.\ {\bf C82}, 02430 (2010).

\bibitem{Chu09}  Y.~Chu,
                 Z.~Ren,
                 T.~Dong, and
                 Z.~Wang,
                 Phys.\ Rev.\ {\bf C79}, 044313 (2009).


\bibitem{arno07} M.~Arnould, 
                 S.~Goriely, and 
                 K.~Takahashi,
                 Phys.\ Rep.\ {\bf 450}, 97 (2007).

\bibitem{paar07} N.~Paar,
                 D.~Vretenar,
                 E.~Khan, and
                 G.~Col\`o,
                 Rep.\ Prog.\ Phys.\ {\bf 70}, 691 (2007).

\bibitem{Grawe07} H. Grawe, K. Langanke, and G. Mart\'{\i}nez Pinedo,
                 Rep.\ Prog.\ Phys.\ {\bf 70}, 1525 (2007).

\bibitem{Jungclaus07} A. Jungclaus et al.,
                 Phys.\ Rev.\ Lett.\ {\bf 99}, 132501 (2007).

\bibitem{Baruah08} S. Baruah et al.,
                 Phys.\ Rev.\ Lett.\ {\bf 101}, 262501 (2008).

\bibitem{Sal05}  F.~Salvat,
                 A.~Jabalonski, and 
                 C.J.~Powell,
                 Comput.\ Phys.\ Commun.\ {\bf 165}, 157 (2005).  

\bibitem{Preston82} M.~A.~Preston and R.~K.~Bhaduri, 
{\it Structure of the Nucleus} (Addison-Wesley, 1982, Reading).

\bibitem{Helm56} R.~H.~Helm, 
                 Phys.\ Rev.\ {\bf 104}, 1466 (1956).

\bibitem{Hof53}  R.~Hofstadter, H.~R.~Fechter, and J.~A.~McIntyre,
                 Phys.\ Rev.\ {\bf 92}, 978 (1953).

\bibitem{Schiff53} L.~I.~Schiff, Phys.\ Rev.\ {\bf 92}, 988 (1953).


\bibitem{G2}     R.~J.~Furnstahl,
                 B.~D.~Serot, and 
                 H.-B.~Tang,
                 Nucl.\ Phys.\ {\bf A615}, 441 (1997); 
                 Nucl.\ Phys.\ {\bf A640}, 505 (1998) (E).

\bibitem{Ser97}  B.~D.~Serot and 
                 J.~D.~Walecka, 
                 Int.\ J.\ Mod.\ Phys.\ {\bf E6}, 515 (1997). 

\bibitem{Arumugam04} P.~Arumugam, 
                 B.~K.~Sharma, 
                 P.~K.~Sahu, 
                 S.~K.~Patra, 
                 Tapas Sil, 
                 M.~Centelles, and 
                 X.~Vi\~nas, 
                 Phys.\ Lett.\ {\bf B601} 51, (2004).

\bibitem{Doba04} J.~Dobaczewski, 
                 M.~V.~Stoitsov, and 
                 W.~Nazarewicz,
                 AIP Conf.\ Proc.\ {\bf 726}, 51 (2004). 

\bibitem{Lala99} G.~A.~Lalazissis,
                 S.~Raman, and
                 P.~Ring,
                 At.\ Data Nucl.\ Data Tables {\bf 71}, 1 (1999).

\bibitem{Est01b} M.~Del Estal,
                 M.~Centelles,
                 X.~Vi\~nas and
                 S.~K.~Patra,
                 Phys.\ Rev.\ {\bf C63}, 044321 (2001).

\bibitem{Ozawa00} A. Ozawa, T. Kobayashi, T. Suzuki, K. Yoshida, and I. Tanihata,
                 Phys.\ Rev.\ Lett.\ {\bf 84}, 5493 (2000).

\bibitem{Otsuka01} T. Otsuka, R. Fujimoto, Y. Utsuno, B. A. Brown, M. Honma, and T. Mizusaki,
                 Phys.\ Rev.\ Lett.\ {\bf 87}, 082502 (2001).

\bibitem{Gupta06} R. K. Gupta, M Balasubramaniam, S. Kumar, S. K. Patra, G. M\"unzenberg,
                 and W. Greiner,
		 J. Phys.\ G: Nucl.\ Part.\ Phys.\ {\bf 32}, 565 (2006).

\bibitem{Tanaka09} K. Tanaka et al.,
                 Phys.\ Rev.\ Lett.\ {\bf 104}, 062701 (2010).

\bibitem{Wang10} N. Wang, M. Liu, and X. Wu,
                 Phys.\ Rev.\ {\bf C81}, 044322 (2010).

\bibitem{Frie82} J.~Friedrich and 
                 N.~Voegler, 
                 Nucl.\ Phys.\ {\bf A373}, 192 (1982). 

\bibitem{Frie86} J.~Friedrich, 
                 N.~Voegler, and 
                 P.-G.~Reinhard, 
                 Nucl.\ Phys.\ {\bf A459}, 10 (1986).

\bibitem{Spru92} D.~C.~Zheng, 
                 N.~Yamanishi, and 
                 D.~W.~L.~Sprung, 
                 Nucl.\ Phys.\ {\bf A550}, 89 (1992).

\bibitem{Est01a} M.~Del Estal, 
                 M.~Centelles, 
                 X.~Vi\~nas, and 
                 S.~K.~Patra, 
                 Phys.\ Rev.\ {\bf C63}, 024314 (2001)

\end{thebibliography}
\end{document}